\documentclass[prb, 11pt]{revtex4-1}
\usepackage{amssymb,amsmath}    
\usepackage{graphicx}   
\usepackage{verbatim}   
\usepackage{color}      
\usepackage{float}      
\begin{document}
\title{Lattice Thermal Conductivity of NiTiSn Half-Heusler Thermoelectric Materials from First-Principles Calculations}
\author{P. Hermet}
\affiliation{Institut Charles Gerhardt Montpellier, UMR-5253, CNRS-Universit\'e de Montpellier-ENSCM, Place E. Bataillon, 34095 Montpellier C\'edex 5, France}
\author{P. Jund}
\affiliation{Institut Charles Gerhardt Montpellier, UMR-5253, CNRS-Universit\'e de Montpellier-ENSCM, Place E. Bataillon, 34095 Montpellier C\'edex 5, France}
\email{pjund@um2.fr}

\begin{abstract}
The microscopic physics behind the lattice thermal conductivity of NiTiSn is investigated using first-principles-based anharmonic lattice dynamics. The calculated lattice thermal conductivity of bulk materials (5.3 W.m$^{-1}$.K$^{-1}$) is in good agreement with the experimental value at the optimal working temperature (700 K), but is overestimated below this temperature. The calculated values can be strongly affected by the size of the crystalline grains. We show that the lattice thermal conductivity is dominated by the acoustic (transverse and mostly longitudinal) modes with no contribution from the optical modes. The acoustic phonons are located below 150~cm$^{-1}$ and involve mainly the tin atoms. The calculated mean free path of the most heat carrying phonons is around fifty nanometers with a maximum life time of $\sim$100~ps. These theoretical results are a step forward in developing the experimental design of low thermal conductivity NiTiSn Heusler based materials.\\
\vspace*{9cm}\\
correponding author: Philippe Jund, pjund@univ-montp2.fr\\
tel: +33 467 149 352, fax: +33 467 144 290
\end{abstract}
\maketitle

\section{INTRODUCTION}
The use of Heusler-type materials in thermoelectric applications (the thermoelectric effect is a direct conversion of a temperature gradient into an electric current via the Seebeck effect) is reinforced by the fact that they are made of environmental friendly elements which explains their current research interest. For a few years, we are interested in Ni-Ti-Sn based Heuslers and in particular in the half-Heusler compound NiTiSn (crystallizing in the F$\bar 4$3m cubic structure \cite{pearson}) which has a maximum thermoelectric efficiency around 700~K \cite{ZT}. This conversion efficiency is measured by the so-called Figure of Merit, $Z=\alpha^2\sigma/\kappa_t$, where $\alpha$ is the thermopower, $\sigma$ is the electrical conductivity and $\kappa_t$ is the total thermal conductivity (lattice and electronic contributions). In previous studies, we have tackled through {\em ab initio} simulations the electronic properties \cite{colinet} and the thermodynamic/mechanical \cite{patNiTiSn,dilato} properties of these compounds with a good agreement with available experimental results. Thus, one aspect of $Z$ still to be addressed is the thermal conductivity. This is even more important since several experimental determinations have shown that it is relatively large ($\approx$ 8~W/m.K at room temperature \cite{expe1,expe2,expe3,expe4}) in NiTiSn and thus needs to be decreased for thermoelectric applications.

One way of decreasing the lattice thermal conductivity (in order to increase $Z$) is to nanostructure the material \cite{expe4} or to include impurities \cite{expe3}. The challenge is also to determine numerically the lattice thermal conductivity to predict the lifetime and the mean free path of the most heat carrying phonons, and thus enlighten the experimentalists for building the most efficient materials. Many theoretical methods have been proposed to solve the Boltzmann transport equations to estimate the lattice thermal conductivity, and they have been summarized in the work of Chernatynskiy and Phillpot \cite{phillpot}.

In the literature, Andrea {\it et al.}~\cite{chaput2} have recently published a value of the calculated lattice thermal conductivity close to $15$~W/m.K \cite{chaput2} at 300~K for NiTiSn in fair agreement with an other {\it ab initio} determination published in 2014 \cite{mingo} finding a value of $17$~W/m.K. A consensus seems thus to exist among the different {\it ab initio} determinations showing a clear overestimation of the lattice thermal conductivity compared to experimental measurements at temperatures below room temperature. This consensus has been put into question by Ding {\it et al.} \cite{ding} who published an {\it ab initio} value of the thermal conductivity of NiTiSn close to $8$~W/m.K in perfect agreement with the experiments. This result poses clearly a problem and one of the objectives of our work is to address this problem taking into account the small amount of calculation details given in Ref. \cite{ding}. 

In this context, we computed the lattice thermal conductivity using the third-order anharmonic lattice dynamics combined with density functional theory calculations. Our method is somewhat different from the direct diagonalization method recently proposed by Chaput \cite{chaput1} since we assume the relaxation time approximation (RTA). In addition, we provide a comprehensive work on the thermal properties of NiTiSn up to 1000~K. In particular, we determine the ideal size of the nanostructures in order to reduce the lattice thermal conductivity, and the phonons and atoms participating most in the heat transport. These original theoretical results should give hints to experimentalists for the design of more efficient Heusler based thermoelectric materials. The influence of the crystallographic grain size on the value of the lattice thermal conductivity is also discussed, leading to a critical discussion of the limits of the present calculations.

\section{THEORETICAL FRAMEWORK}
 
	\subsection{Lattice thermal conductivity} 
 
The lattice thermal conductivity, $\kappa$, has been obtained using the Boltzmann transport equation within the relaxation time approximation of the phonon gas. For an isotropic system, it is given by~\cite{Srivastava}:
\begin{equation}
\label{eqKappa}
\kappa = \frac{1}{3\Omega N}\sum_{\mathbf{q},s} C_{\mathbf{q}s} v^2_{\mathbf{q}s} \tau_{\mathbf{q}s},
\end{equation}
where $C_{\mathbf{q}s}$ is the lattice specific heat of the phonon with wave-vector $\mathbf{q}$ and branch $s$, $v$ is the average group velocity, $N$ is the number of $q$-points and $\Omega$ is the volume of the primitive unit cell. The relaxation time, $\tau$, represents the time after which a phonon reaches equilibrium on the average and depends on the scattering processes involved. In a pure bulk sample, the only source of phonon scattering is anharmonicity which is usually dominated by the three-phonon processes. Using the perturbation theory,  the expression of the relaxation time becomes~\cite{Maradudin,Cowley}:
\begin{equation}
\tau_{\mathbf{q}s}=\frac{1}{2\Gamma_{\mathbf{q}s}},
\end{equation}
where
\begin{equation} \begin{split}
\Gamma_{\mathbf{q}s}=&\frac{\pi\hbar}{16N}\sum_{\mathbf{q'},s'}\sum_{\mathbf{q''},s''} \frac{|\mathbf{A}(\mathbf{q}s,\mathbf{q'}s',\mathbf{q''}s'')|^2}{\omega_{\mathbf{q}s}\omega_{\mathbf{q'}s'}\omega_{\mathbf{q''}s''}}\Big\{ (n_{\mathbf{q'}s'}+n_{\mathbf{q''}s''}+1)\delta(\omega_{\mathbf{q}s}-\omega_{\mathbf{q'}s'}-\omega_{\mathbf{q''}s''})+\\
&(n_{\mathbf{q'}s'}-n_{\mathbf{q''}s''})[\delta(\omega_{\mathbf{q}s}+\omega_{\mathbf{q'}s'}-\omega_{\mathbf{q''}s''})-\delta(\omega_{\mathbf{q}s}-\omega_{\mathbf{q'}s'}+\omega_{\mathbf{q''}s''})]\Big\}.
\end{split} \end{equation} 
In these equations, $\hbar$, $\omega$ and $n$ are the reduced Planck constant, frequency, and Bose-Einstein distribution of phonons, respectively. The delta function imposes the energy conservation of the three-phonon scattering process. The three-phonon matrix element is given by:
\begin{equation} \begin{split}
\mathbf{A}(\mathbf{q}s,\mathbf{q'}s',\mathbf{q''}s'')= &\sum_{j,l=0}\ \sum_{j',l'}\ \sum_{j'',l''}\ \sum_{\alpha,\beta,\gamma}\phi_{\alpha\beta\gamma}(j 0,j' l',j'' l'') \frac{e_{\alpha j}(\mathbf{q}s)e_{\beta j'}(\mathbf{q'}s')e_{\gamma j''}(\mathbf{q''}s'')}{\sqrt{M_j M_{j'} M_{j''}}} \\
&e^{i\mathbf{q'.R_{l'}}+i\mathbf{q''.R_{l''}}}\ \delta_{\mathbf{q}+\mathbf{q'}+\mathbf{q''},\mathbf{G}},
\end{split} \end{equation} 
where $\alpha$, $\beta$, and $\gamma$ label the Cartesian directions, $\phi$ are the third-order force constants, $M$ are the atomic masses, $e_{\alpha j}(\mathbf{q}s)$ is the polarization vector of atom $j$ along the $\alpha$-direction which corresponds to phonon $\mathbf{q}s$, and $\mathbf{R}_l$ is the position vector of the $l$-th cell. The delta function describes the momentum conservation during the normal ($\mathbf{G}=0$) and umklapp ($\mathbf{G}\ne0$) processes, where $\mathbf{G}$ is a reciprocal lattice vector. 

The boundary effect in the lattice thermal conductivity has been considered using the Matthiessen rule:
\begin{equation}
\frac{1}{\tau_{\mathbf{q}s}}=\frac{1}{\tau_{\mathbf{q}s}^{ph-ph}}+\frac{2v_{\mathbf{q}s}}{L},
\end{equation}
where the first term is the scattering rate due to phonon-phonon interactions (Eq. 1) while the second term is the scattering rate due to the grain boundaries of size $L$. In this formalism, the processes that scatter phonons are assumed independent and described by individual scattering rates. The point-defect scattering that can be present in the NiTiSn crystals is not considered in our work.

\subsection{Computational details} 

Structure relaxation and Hellmann-Feynman forces are determined within the density functional theory (DFT) framework as implemented in the VASP package~\cite{VASP}, and the generalized gradient approximation (GGA) for the exchange correlation functional as proposed by Perdew, Burke and Ernzerhof \cite{PBE}. The interactions between ions and electrons are described by the projector augmented wave method \cite{PAW} in the real space representation. Ni($3d^{8}$, $4s^2$), Ti($3d^{2}$, $4s^2$), and Sn($5s^{2}$, $5p^2$)-electrons are considered as valence states. The electronic wave functions are expanded in plane-waves up to a kinetic energy cutoff of 425~eV and integrals over the Brillouin zone are approximated by sums over a 4$\times$4$\times$4 mesh of special $k$-points according to the Monkhorst-Pack scheme~\cite{Monkhorst}. 

The third-order force constants, $\phi$, are obtained from the finite displacement method on a 2$\times$2$\times$2 supercell (96 atoms) built from the conventional cell. All the interactions inside this supercell are considered, which represents 13 nearest neighbors. In this context, we have displaced by 0.03~\AA\ one or two atoms at a time from their equilibrium positions along the Cartesian coordinates, leading to a total of 3 single-point energy calculations for the harmonic part and 301 single-point energy calculations for the anharmonic part. The phonon frequency and the polarization vectors are obtained straightforwardly by diagonalizing the analytical dynamical matrix (transverse optical phonon modes). Because NiTiSn is a polar semiconductor, we have also computed the non-analytical dynamical matrix to include longitudinal optical phonon modes (see Refs. \cite{pat1,pat2} for more information). Born effective charges and electronic dielectric tensors are calculated by density functional perturbation theory.

Numerically, the lattice thermal conductivity, $\kappa$, is calculated using the tetrahedron method over a 20$^3$ mesh of $q$-points. This grid gives converged results since a 30$^3$
$q$-mesh leads to an increase of $\kappa$ by only 1.5\% at 300~K.

\section{RESULTS AND DISCUSSION}

The calculated lattice thermal conductivity of NiTiSn is represented in Figure 1 as a function of temperature for a bulk sample using Eq.~1 and taking into account the grain size of a polycrystalline sample following Eq.~5. These results are compared with two of the available experimental results \cite{expe1,expe2} covering the whole temperature range. 

\begin{figure}[H]
\begin{center}
\includegraphics[width=10cm]{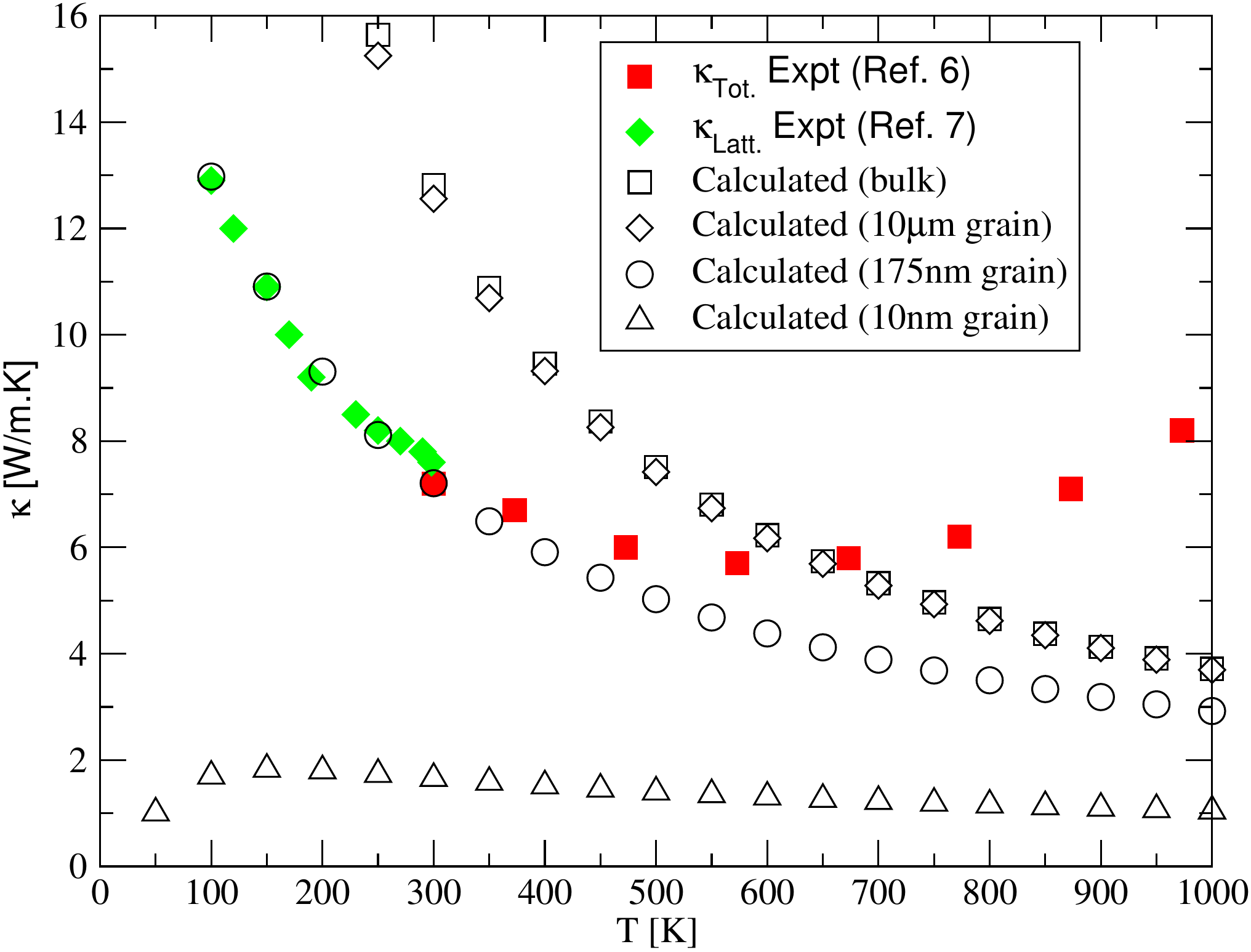}
\caption{Temperature dependence of the calculated lattice thermal conductivity of bulk NiTiSn along with experimental data reported in the literature. Three different NiTiSn grain sizes (10, 175 and 10000 nm) have also been considered in our calculations.}
\label{Figure 1}
\end{center}
\end{figure}
The first ascertainment is that the calculated bulk results are overestimated compared to the experimental values up to 600~K. Above this temperature, the experimental total thermal conductivity determined in Ref. \cite{expe1} increases with temperature because of the bipolar contribution (electron-hole pairs) to the electronic thermal conductivity, contribution that is not taken into account in our calculations. This electronic contribution seems to be negligible at low temperatures as demonstrated by the perfect connection between the two experimental curves (in Ref. \cite{expe2} only the lattice thermal conductivity is measured). In addition, we have shown in a previous work that even at high temperatures the electronic contribution to the heat capacity and the thermal expansion coefficient is small \cite{dilato}. At the optimum working temperature close to 700~K \cite{ZT}, the predicted $\kappa$ is correct but this agreement may be fortuitous. At lower temperatures and especially at 300~K, our calculated value ($\approx$ 12.8~W/m.K) is clearly overestimated compared to the experimental value ($\approx$ 8~W/m.K), which is also the case for other recent theoretical determinations \cite{chaput2,mingo} as stated earlier. This disagreement with experiments, already noticed for the calculation of the thermal expansion \cite{dilato}, can have several origins.

On the one hand, theoretically, it can be connected to the size of the supercell considered in our calculations or to a number of interacting neighbors that is not large enough. In our calculations, we have considered all the interactions inside a 2$\times$2$\times$2 supercell, which represents 13 nearest neighbors. This is in contrast with the paper of Ding {\it et al}~\cite{ding} in which the authors only took into account 3rd nearest neighbors \cite{ding-private}. Thus, we also performed our calculations limiting the number of nearest neighbors to 3 and found an increase of $\kappa$ to 15.2~W/m.K. Ding {\it et al.} considered a 4x4x4 supercell of the primitive cell containing 192 atoms (so twice more than the present study) which could explain the value of 7.6~W/m.K they obtained for $\kappa$ at 300~K. However, this is in contradiction with the value of 17.9~W/m.K obtained with the {\it same} method of calculation and the {\it same} size of the supercell by Carrete {\it et al.} \cite{mingo}. It is worth noting that these last authors are also the developers of the program used by Ding {\it et al.} namely ShengBTE \cite{sbte}. At last since it is not specified in the paper of Ding {\it et al.}~\cite{ding}, we also performed the DFT calculations at the LDA level (instead of the GGA) and with the lattice parameter used in their work (5.89~$\AA$), but again we found a thermal conductivity of 16.9~W/m.K. In short, we are unable to reproduce the results of Ding et al. or to explain why these authors found a value of $\kappa$ so small. Until this exception is reproduced and explained, our value of $\approx$ 12.8~W/m.K is the closest to the experimental value among all the (coherent) DFT-based determinations of $\kappa$ even though it represents an overestimation by 50\% of the measured value.

On the other hand, experimentally, this disagreement could be due to the quality of the experimental samples that are often polycrystalline, or polyphasic, or contain impurities as highlighted by the different experimental results below room temperature~\cite{expe1,expe2,expe3,expe4}. It is thus not surprising to find an overestimated value of $\kappa$ for calculations performed on a perfect monocrystal. To assess the polycrystalline nature of the experimental samples, we show in Figure~1 the calculated $\kappa$ for three different grain sizes. At low temperatures, the agreement with experiments is best for a grain size of the order of 175~nm. A grain size of 10~nm is clearly too small since the thermal conductivity is underestimated, whereas for a grain size of 10~$\mu$m the bulk thermal conductivity is recovered which is coherent with the experimental findings of Bhattacharya {\it et al.}~\cite{expe2}. 
Nevertheless the grain size alone can not explain the disagreement between experiments and simulations since the same authors find a value of $\approx$ 8~W/m.K for a grain size of $\approx$ 8~$\mu$m ~\cite{expe2}. We have previously shown that point defects are very likely to form in NiTiSn and especially interstitial Ni atoms~\cite{colinet}. These defects are suspected to be at the origin of the measured electronic bandgap which is lower than the calculated GGA bandgap obtained for pure NiTiSn~\cite{colinet}, a situation that is quite unusual. These impurities, but also other kinds of defects (local disorder, voids in the structure, multiple phases,...) that are not considered in our calculations, could explain why the experimental thermal conductivity is significantly lower than the calculated one according to the usual Matthiessen type models~\cite{holland}.      

In the following, we will limit our temperature analysis of the lattice thermal conductivity to two temperatures: room temperature and 700~K (the optimum working temperature). The calculated contribution of the acoustic and optical phonon frequencies to the lattice thermal conductivity is reported in Figure 2. 
\begin{figure}[H]
\begin{center}
\includegraphics[width=10cm]{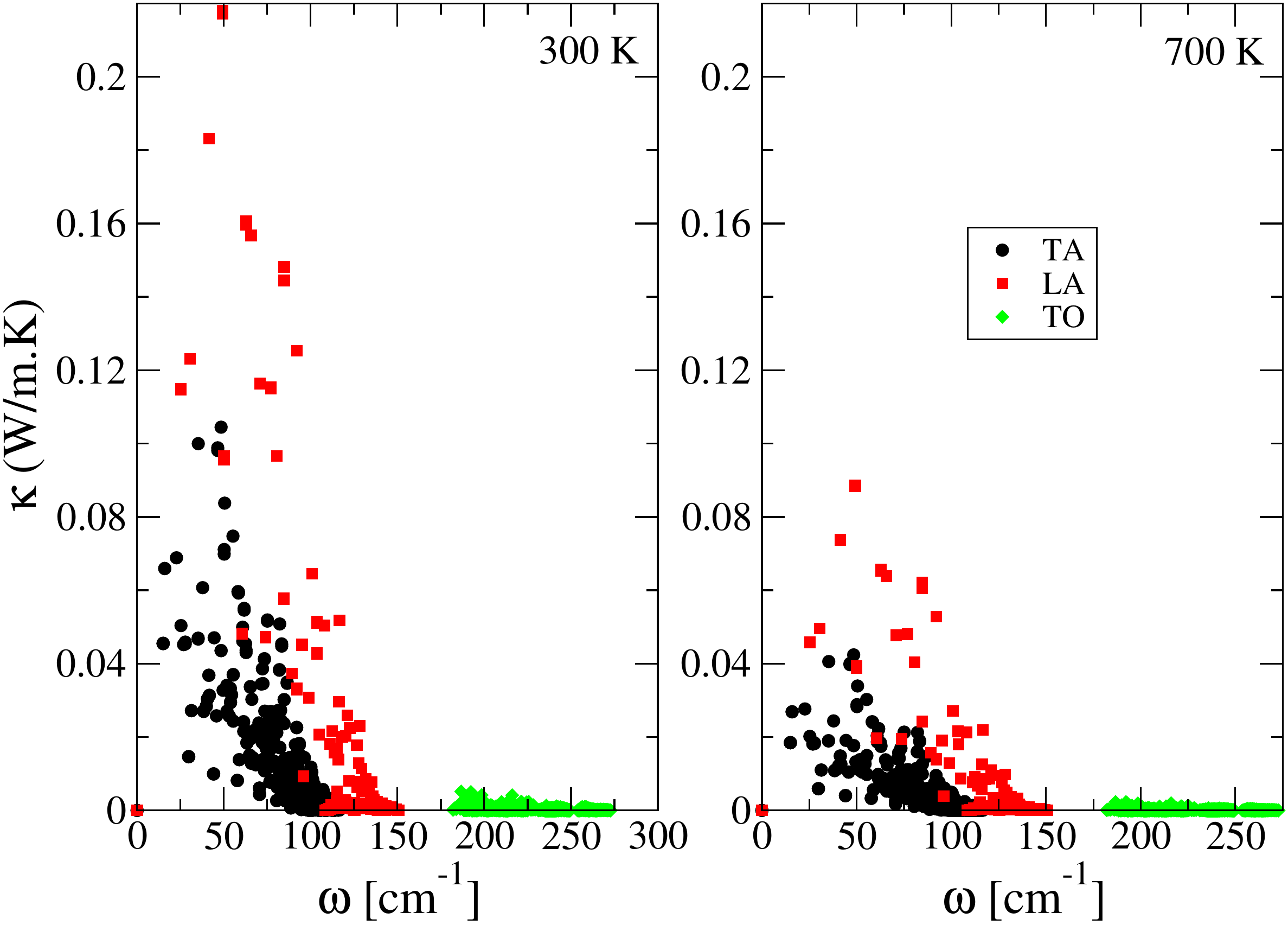}
\caption{Calculated contribution of the acoustic and optical phonon modes to the lattice thermal conductivity at 300 and 700~K.}
\end{center}
\end{figure}
We observe that only the acoustic phonons (transverse and longitudinal) with frequencies below 150~cm$^{-1}$ dominate the lattice thermal conductivity regardless of the temperature similarly to what we observed in chalcogenide glasses\cite{chalco}. This explains why the elementary model proposed by Slack~\cite{slack}, in which only the phonon in the acoustic branches carry the heat, is reliable for NiTiSn. Indeed, using a Gr\"uneisen parameter of 1.60 and a Debye temperature of 360~K from Ref.\cite{dilato}, the Slack model gives $\kappa^{Slack}$ = 14~W/m.K at 300~K which is only slightly higher than our calculated value (12.8~W/m.K). The atom-projected density-of-states (Figure 3) shows that the acoustic phonons with a frequency lower than 150 cm$^{-1}$ mainly involve tin atoms. 

\begin{figure}[H]
\begin{center}
\includegraphics[width=10cm]{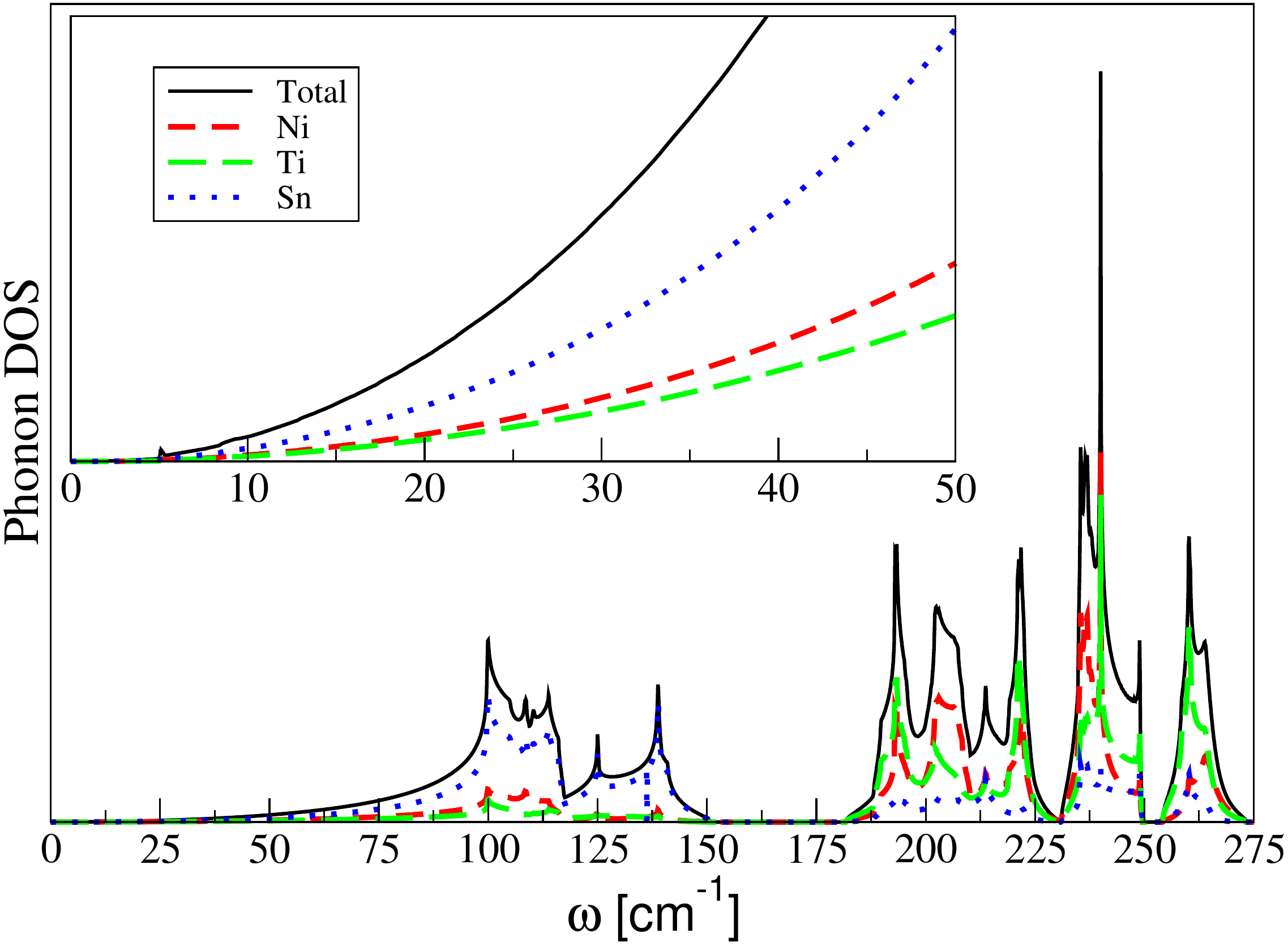}
\caption{Total and atom-projected density-of-states of NiTiSn. Inset: zoom of the 0-50 cm$^{-1}$ range.}
\end{center}
\end{figure}
Thus, substitutions on the Sn-site should be undertaken to lower the lattice thermal conductivity. Experimentally Sb substitutions have been undertaken on the Sn site \cite{expe2} but the small mass difference and the concomitant increase of the grain size didn't permit to observe a decrease of $\kappa$. To reach the low values necessary for thermoelectric applications, the dimensions of the sample can be decreased as recently shown by Ho{\l}uj {\it et al.}~\cite{NiTiSnlayers}, or the sample can be nanostructured. Concerning this last technique, the natural question that arises immediately is: what is the optimal size of the nanostructures ?


To answer this question, we show in Figure~4 the relaxation time and the mean free path of the phonons. 

\begin{figure}[H]
\begin{center}
\includegraphics[width=10cm]{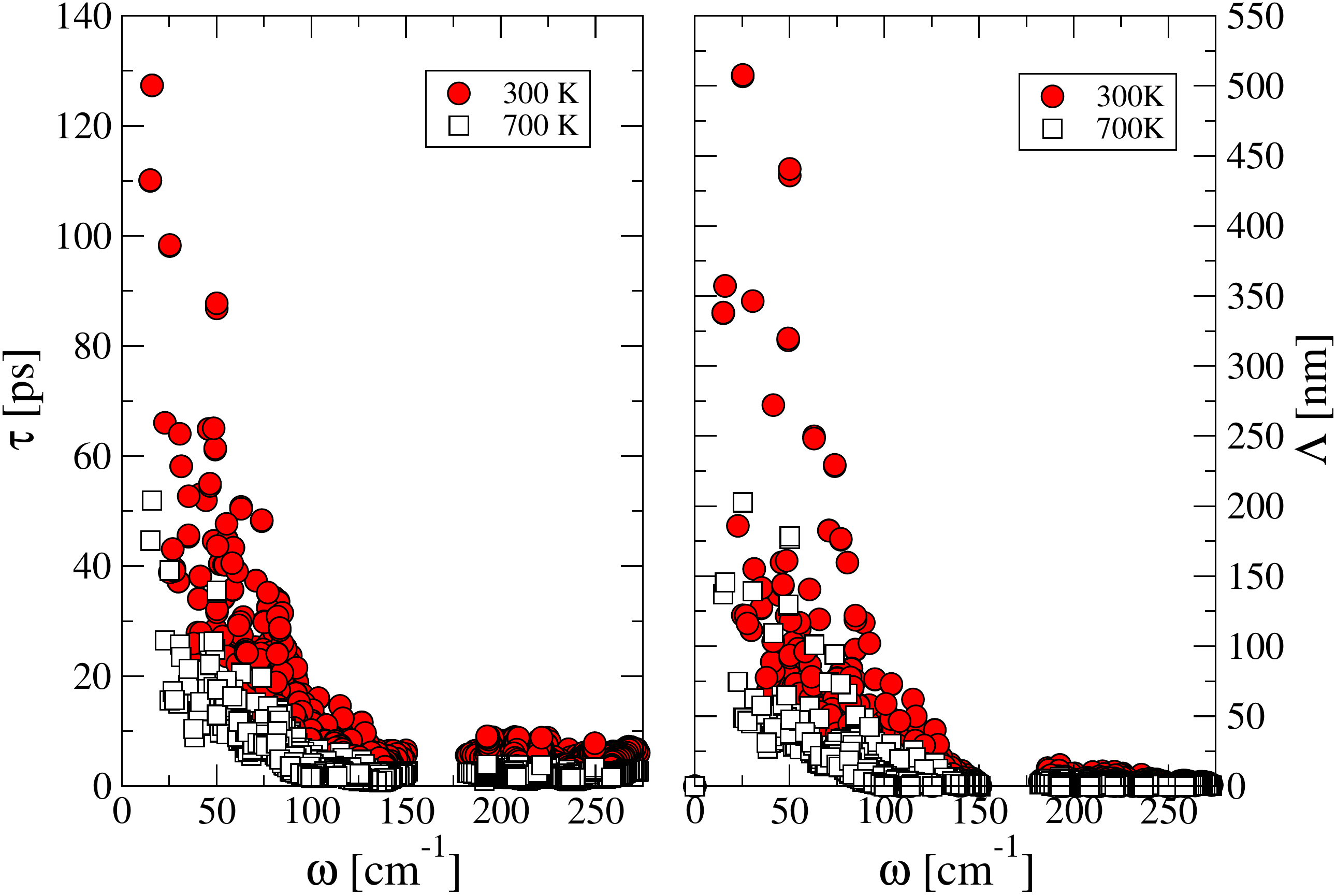}
\caption{Frequency-dependent relaxation time ($\tau$) and mean free path ($\Lambda$) of phonons calculated at 300 and 700~K.}
\end{center}
\end{figure}
At 300~K, the relaxation time of the low frequency acoustic phonons is around 100~ps, a value divided by two at 700~K (Figure 4, left). Concerning the mean free path of the phonons (Figure 4, right), phonons with a frequency close to 25 cm$^{-1}$ travel as far as 500~nm at 300~K, a distance that is reduced to 200~nm at 700~K. Thus some of the acoustic phonons travel far but it is interesting to check if they contribute greatly to $\kappa$ in order to predict a sensible size of the experimental nanostructuration. The contributions of the acoustic phonons to the lattice thermal conductivity as a function of their mean free path are represented at the 300~K (Figure 5, left) and 700~K (Figure 5, right). The two parts of Figure~5 are very similar indicating that the effect of temperature is mainly a reduction of the mean free path of the phonons when going from 300 to 700~K. This reduction is roughly equal to the ratio of the two temperatures (700/300) which is a direct consequence of the 1/T variation of the mean free path at high temperature. Globally speaking the phonons that contribute more than 1\% to $\kappa$ are longitudinal phonons with mean free paths smaller than or close to half of the maximum mean free path. It is worth noting that the phonons with the largest $\Lambda$ contribute more than 0.5\% to $\kappa$.      

\begin{figure}[H]
\begin{center}
\includegraphics[width=10cm]{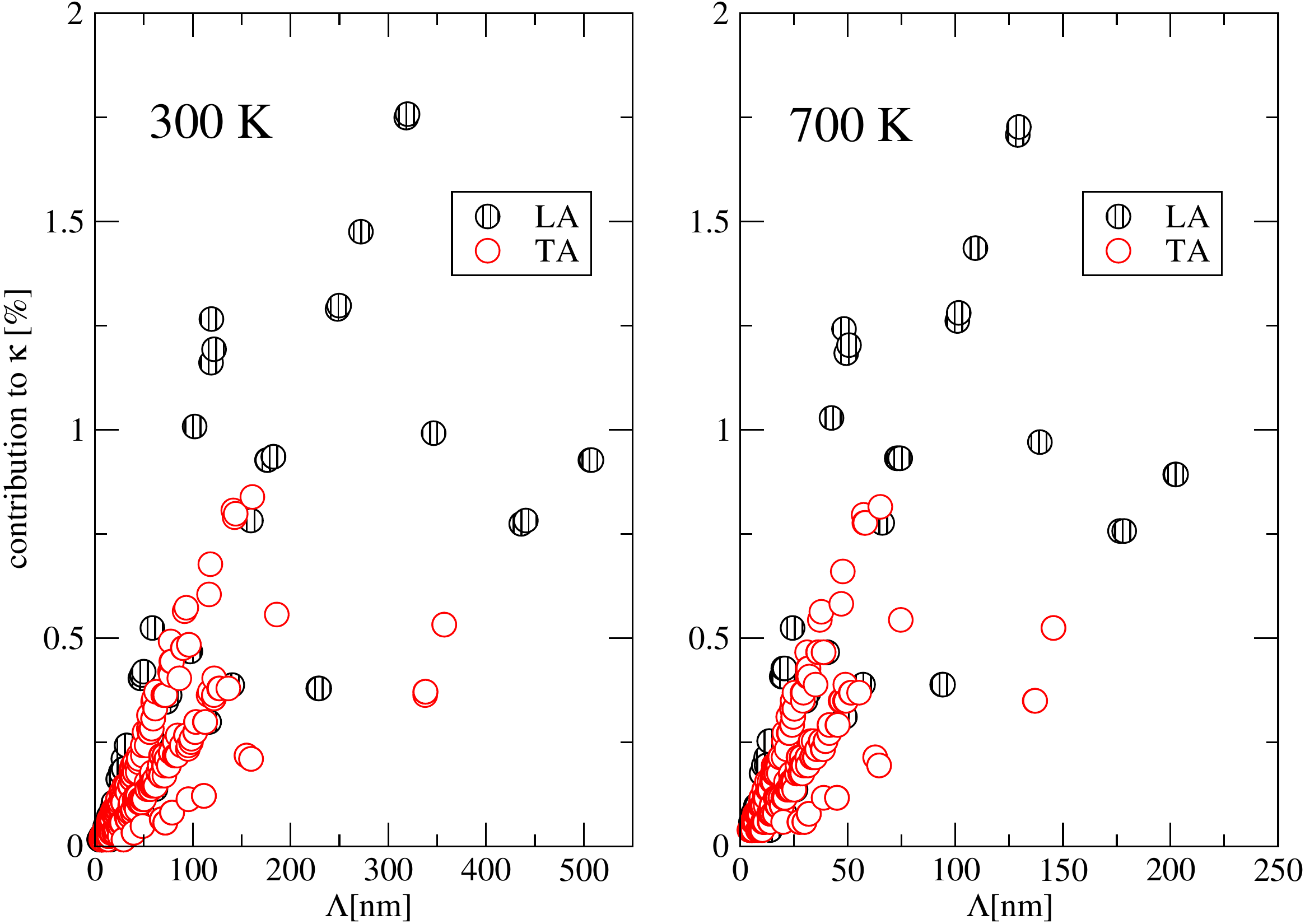}
\caption{Contribution of the acoustic phonons (transverse and longitudinal) to $\kappa$ as a function of their mean free path ($\Lambda$) at 300 and 700~K.}
\end{center}
\end{figure}

Nevertheless these phonons are not numerous, so one has to consider the sum of all these contributions to have an idea of the ideal size of the nanostructures to sensitively diminish the thermal conductivity. This can be achieved by determining the cumulative thermal conductivity ($\kappa_C$) \cite{cumul} which is shown in Figure 6 as a function of $\Lambda$ at 300 and 700~K.   

\begin{figure}[H]
\begin{center}
\includegraphics[width=10cm]{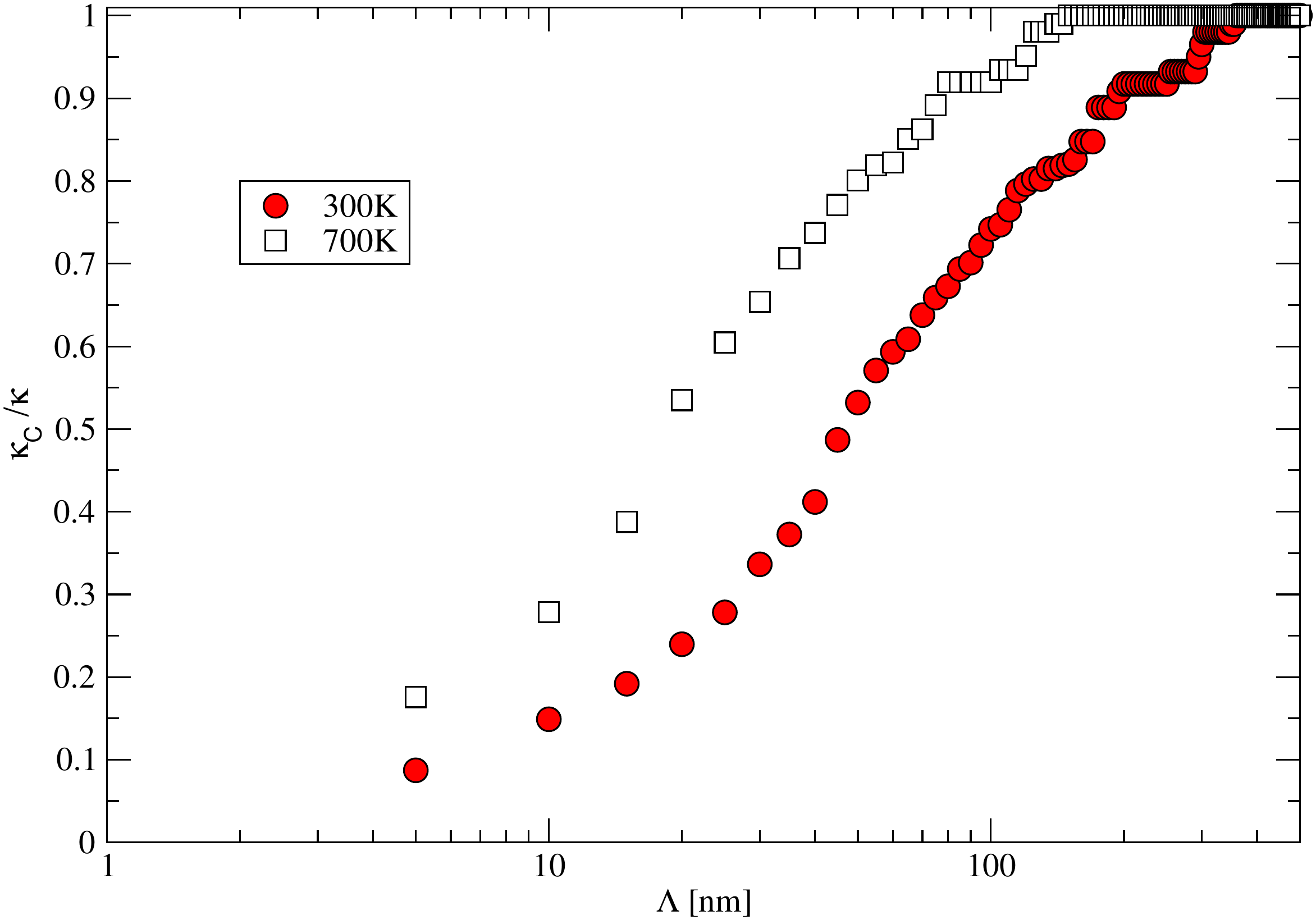}
\caption{Cumulative thermal conductivity $\kappa_C$ normalized by the bulk value $\kappa$ versus the phonons mean free path ($\Lambda$) at 300 and 700~K.}
\end{center}
\end{figure}

This quantity accumulates the contribution to $\kappa$ from all the phonons with mean free paths smaller than $\Lambda$. The first information that can be extracted from Figure 6 is that the curves reach 1 in the long mean free path limit indicating that our wavevector mesh used to perform the calculations is fine enough. The second and most interesting information is that phonons with $\Lambda$s up to $\approx$ 50~nm ($\approx$ 20~nm) represent 50\% of $\kappa$ at 300~K (700~K). Thus, nanostrucutures of this size can contribute to annihilate phonons with 50\% contribution to $\kappa$ since the phonons with mean free paths larger than this size become ballistic and will be inhibited by the size effect \cite{size}. These calculated values of the optimal size of the nanostructures are probably the upper bounds since our calculated bulk lattice thermal conductivity is overestimated with respect to the experiment especially at 300~K. Nevertheless, this theoretical result is an important technological information for experimentalists who wish to design nanostructured half-Heusler samples exhibiting weak thermal conductivity for thermoelectric applications.

\section{CONCLUSIONS}

The lattice contribution to the thermal conductivity of NiTiSn half-Heusler compounds has been calculated using the third-order anharmonic lattice dynamics combined with density functional theory calculations. First, the calculated bulk values are overestimated at 300~K compared to the different experimental values, but are close at 700~K which is the optimal working temperature for these thermoelectric materials. Nevertheless, our calculations show that these values can be strongly affected by the size of the grains (like in polycrystalline samples), probably because the lattice thermal conductivity of bulk (monocrystalline) NiTiSn is high. Then, we find that only the acoustic (transverse and mostly longitudinal) phonons with frequencies below 150~cm$^{-1}$ dominate the thermal conductivity. The atom-projected density-of-states shows that these phonons mainly involve tin atoms. Finally, we have determined the relaxation time and the mean free path of the heat carrying phonons and propose that the ideal size for nanostructuring NiTiSn is around fifty nanometers. This result gives the upper limit to the experimental design of low thermal conductivity NiTiSn Heusler based materials.










\clearpage
Graphical abstract:

\begin{figure}
\begin{center}
\includegraphics[width=8.5cm, height=4.75cm]{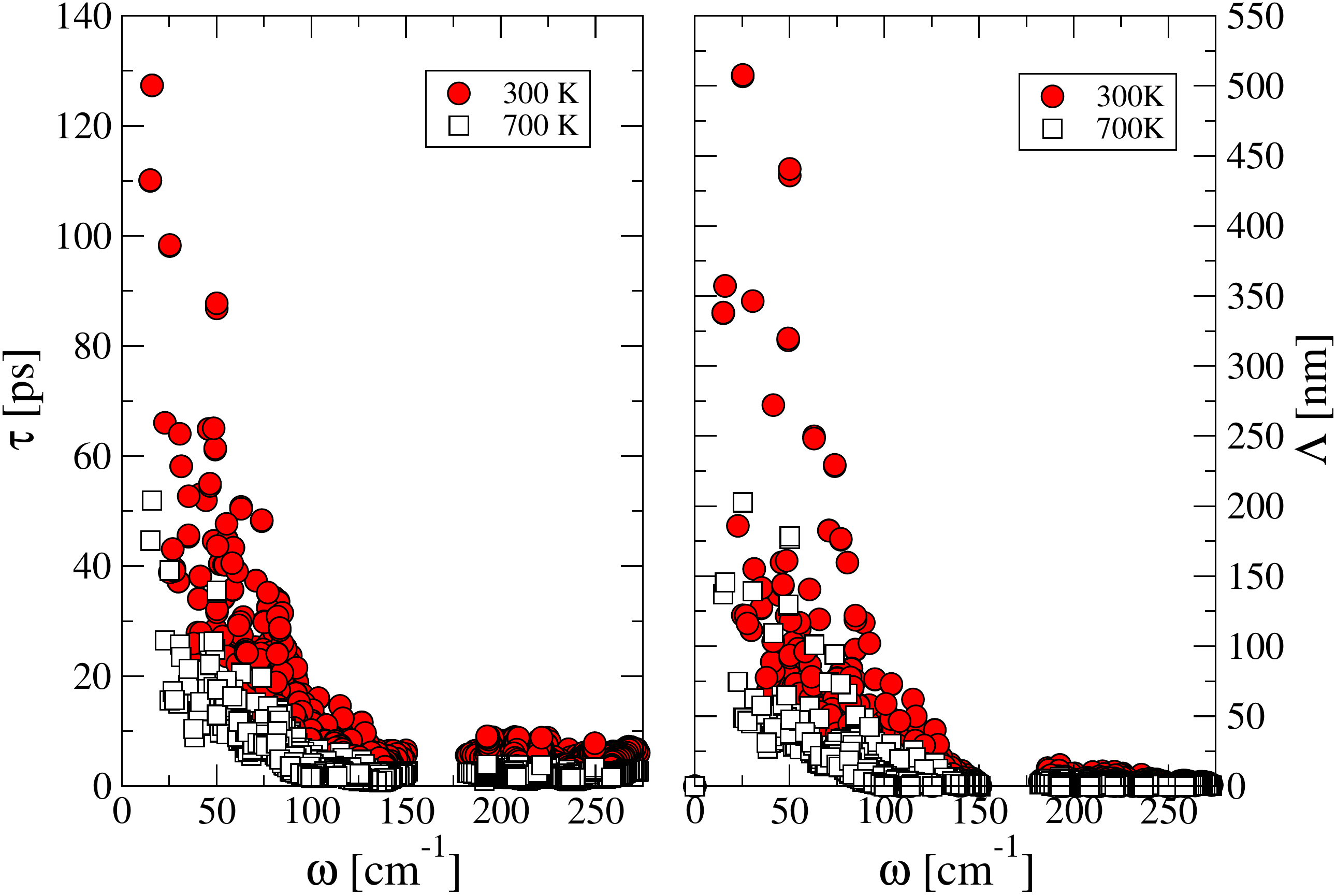}
\end{center}
\end{figure}

\end{document}